# Optical Nanoimaging of Hyperbolic Surface Polaritons at the Edges of van der Waals Materials

*P. Li[1,\*], I. Dolado[1,\*], F. J. Alfaro-Mozaz[1], A. Y. Nikitin[1,2], F. Casanova[1], L. E. Hueso[1], S. Vélez[1], and R. Hillenbrand[2,3,#]*

[1]CIC nanoGUNE, E-20018, Donostia-San Sebastián, Spain.
[2]IKERBASQUE, Basque Foundation for Science, 48011 Bilbao, Spain.
[3]CIC NanoGUNE and UPV/ EHU, E-20018, Donostia-San Sebastian, Spain

[\*]Equally contributing authors
[#]Correspondence to: r.hillenbrand@nanogune.eu

**Hyperbolic polaritons in van der Waals (vdW) materials recently attract a lot of attention, owing to their strong electromagnetic field confinement, ultraslow group velocities and long lifetimes. Typically, volume-confined hyperbolic polaritons (HPs) are studied. Here we show the first near-field optical images of hyperbolic surface polaritons (HSPs), which are confined and guided at the edges of thin flakes of a vdW material. To that end, we applied scattering-type scanning near-field optical microscopy (s-SNOM) for launching and real-space nanoimaging of hyperbolic surface phonon polariton modes on a hexagonal boron nitride (h-BN) flake. Our imaging data reveal that the fundamental HSP mode exhibits stronger field confinement (shorter wavelength), smaller group velocities and nearly identical lifetimes, as compared to the fundamental HP mode of the same h-BN flake. Our experimental data, corroborated by theory, establish a solid basis for future studies and applications of HPs and HSPs in vdW materials.**

Polaritons –electromagnetic fields coupled to phonons or plasmons– in van der Waals (vdW) materials can naturally exhibit a hyperbolic dispersion, as in certain frequency regions the out-of-plane (axial) permittivity can be opposite in sign to the in-plane (transverse) permittivity.[1-19] These so-called hyperbolic polaritons (HPs) can propagate with arbitrarily large wavevectors **k** through the material, thus enabling subdiffraction focusing and waveguiding[5,6,9,10] or negative refraction[11]. They have manifold nanophotonic applications, including super-resolution optical imaging[6] or subwavelength-sized resonators[7-8], among others. The required thin flakes of high crystal quality can be readily obtained by standard exfoliation techniques. The possibility to obtain one-atom-thick flakes further promises the development of atomic-scale hybrids for visible, mid-infrared and terahertz nanophotonics.[20]

At surfaces exhibiting a strong in-plane anisotropy, interestingly, vdW materials can also support hyperbolic surface polaritons[16,21-25] (HSPs, also called Dyakonov surface waves[26]). In contrast to HPs, they are confined to surfaces and interfaces, similar to conventional surface polaritons[27]. HSPs in vdW materials have been studied mostly theoretically.[16,23-25] Only recently, hyperbolic surface plasmon polaritons were observed experimentally by electron energy loss spectroscopy in $Bi_2Se_3$ platelets.[13]

HSP also exist in artificial metamaterials and on artificial metasurfaces.[21-23,26,28] At visible frequencies, the extraordinary propagation of HSPs on a silver/air grating has been recently demonstrated by far-field fluorescence imaging.[28] The unit cells or layers of artificial metamaterials and metasurfaces are often obtained by lithographic structuring, leading to typical structure sizes $S$ in the range of several 1 to 10 nanometers. The maximum HSP wavevector is thus limited according to $k_{HSP} \sim 1/S$. vdW materials offer the advantage that the HSP wavevector $k_{HSP}$ can reach its ultimate maximum, because of the atomic-scale distances between the atomically thick layers.

Here we perform the first comparative theoretical and experimental study of volume and surface-confined hyperbolic phonon polaritons in h-BN flakes. Being a typical insulating and polar vdW material, h-BN exhibits hyperbolic phonon polaritons in two mid-infrared spectral regions: (1) the lower Reststrahlen band (760-825 cm$^{-1}$) with a negative out-of-plane permittivity ($\varepsilon_{\parallel}<0$) and positive in-plane permittivity ($\varepsilon_{\perp}>0$); (2) the upper Reststrahlen band (1360-1610 cm$^{-1}$) with $\varepsilon_{\perp}<0$ and $\varepsilon_{\parallel}>0$.[5-10] In this work, we demonstrate HPs and HSPs by numerical simulations and discuss how they can be excited by a dipole source. For experimental real-space imaging of HPs and HSPs, we apply scattering-type scanning near-field optical microscopy[5,6,8-10] (s-SNOM). The images reveal that HSPs propagate along the edges (sidewalls) of thin h-BN flakes, which form a natural two-dimensional (2D) waveguide for HSPs. By spectroscopic s-SNOM imaging, we measure the dispersion relation, spatial confinement, group velocities and lifetimes of the fundamental HSP waveguide mode at the edge of the h-BN flake, and compare them with that of the fundamental HP waveguide mode measured for the same flake. The experimental results are corroborated by full-wave simulations.

Hyperbolic polaritons can be excited with a dipole source, which generates highly concentrated near fields that provide wavevectors larger than that of free-space photons. In Fig. 1a we demonstrate the launching of HPs in h-BN with the help of numerical simulations. To that end, we place an electric dipole (oriented vertically) on top of bulk h-BN (infinitely thick), whose optical axis (OA) is oriented perpendicular to the horizontal surface (illustrated in Fig. 1a, left). We observe highly confined bright rays, which propagate inside the h-BN, revealing volume-confined hyperbolic phonon polaritons (Fig. 1a, right). As the HP dispersion relation does not depend on the direction of the in-plane wavevector, the HP rays launched by a vertically oriented dipole form a three-dimensional (3D) cone[5-10], which we illustrate by virtual cuts through the h-BN (Fig. 1a, right). We stress that the dipole does not excite hyperbolic surface polaritons, which is apparent from the absence of any field confinement at the h-BN surface.

As introduced before, the excitation of HSPs on h-BN requires a strong in-plane anisotropy of the surface[21-25,28]. It can be achieved when the atomic h-BN layers are perpendicular to the surface (i.e. when the OA is parallel to the surface, as sketched in Fig.1b), which we demonstrate by rotating the h-BN by 90 degrees as compared to Fig. 1a. Within the surface, the permittivity parallel to the OA is positive, while the one perpendicular to the OA is negative. In the numerical simulations (Fig. 1b, right) we consequently see bright rays that propagate along the h-BN surface in specific directions (neither parallel nor perpendicular to the OA). Most important, the fields associated with the rays are confined to the surface, corroborating that HSPs are observed. We note that the dipole also launches HPs, which manifest as rays that propagate inside the material. In contrast to Fig. 1a, they do not form a cone but rather propagate in specific directions (see right virtual cut through the h-BN in Fig. 1b).

In order to understand the highly directional propagation of HP and HSP rays, we analyze their wavevectors $\boldsymbol{k} = (k_{\perp,a}, k_{\perp,b}, k_{\parallel})$, which are given by the solutions of the following dispersion relations, respectively:[1-10]

$$\frac{k_{\perp,a}^2 + k_{\perp,b}^2}{\varepsilon_{\parallel}} + \frac{k_{\parallel}^2}{\varepsilon_{\perp}} = k_0^2 \qquad (\text{HPs}^{1-10}) \qquad (1a)$$

$$\frac{k_{\perp,a}^2 - k_e^2}{\varepsilon_{\parallel}} + \frac{k_{\parallel}^2}{\varepsilon_{\perp}} = k_0^2 \qquad (\text{HSPs}^{22,26}) \qquad (1b)$$

$k_{\parallel}$ is the wavevector parallel to the OA and $k_{\perp}$ are the wavevectors perpendicular to the OA. $k_0$ is the free-space wavevector. $k_e$ is the HSP wavevector (in the h-BN) perpendicular to the surface[22], which can be simplified to $k_e \approx -(k_{\perp,a}^2 + k_{\parallel}^2)^{1/2} \varepsilon_d / \varepsilon_{\perp}$ for large wavevectors (see Note 1 in the Supporting Information). The solutions of Eq. 1a for a given frequency are three-dimensional open hyperboloids (isofrequency surfaces, Fig. 1c). They allow extremely large HP wavevectors ($k_{HP} >> k_0$) to exist, which can be described by an asymptote exhibiting an angle $\theta_v$ relative to $k_{\perp}$. A dipole field close to the surface – comprising wavevectors much larger than $k_0$ – thus leads to a highly directional launching and propagation of HPs, which is determined by $\theta_v$. The three-dimensional HP propagation (Fig. 1a) is thus described by a cone of opening angle $2\theta_v$ (Fig. 1c). The solutions of Eq. 1b are two-dimensional hyperbola (isofrequency curves, red curve in Fig. 1d). Similar to HPs, we find that the large HSP wavevectors $k_{HSP}$ can be described by an asymptote with angle $\theta_s$ relative to $k_{\perp,a}$. Excitation of HSPs by a dipole field thus leads to highly directional two-dimensional HSP propagation along the surface (Fig. 1b).

To compare HSPs and HPs, we plot in Fig. 1d a cross-section of the HP isofrequency surface (blue curve) extracted from Fig. 1c additional to the HSP isofrequency line (red curve). Although the shapes of the two curves are similar, the directions of their large-wavevector asymptotes are different. The angles $\theta_v$ and $\theta_s$ are given by

$$\theta_v = \arctan\sqrt{-\varepsilon_{\perp}/\varepsilon_{\parallel}} \qquad (\text{HPs}^{5-10}) \qquad (2a)$$

and

$$\theta_s = \frac{\pi}{2} - \arcsin\sqrt{\frac{\varepsilon_d^2 + |\varepsilon_{\perp}|\varepsilon_{\parallel}}{|\varepsilon_{\perp}|(\varepsilon_{\parallel} + |\varepsilon_{\perp}|)}} \qquad (\text{HSPs}^{22}) \qquad (2b)$$

where $\varepsilon_d$ is the permittivity of the adjacent isotropic medium. In contrast to HPs, the propagation of HSPs is not only dependent on the permittivity (and thus frequency) of the hyperbolic materials, but also on $\varepsilon_d$. In Fig. 1e we compare the calculated angles $\theta_s$ and $\theta_v$ for h-BN HPs within the upper Reststrahlen band. It is clearly seen that the HP propagation angle exhibits a maximum $\theta_v$ at the transverse optical (TO) phonon frequency $\omega$=1360 cm$^{-1}$ of h-BN and a minimum $\theta_v$ at the longitudinal optical (LO) phonon frequency $\omega$=1610 cm$^{-1}$. However, the minimum $\theta_s$ of HSP propagation is reached at the surface phonon optical (SO) frequency $\omega$=1568 cm$^{-1}$, which is determined by $\varepsilon_\perp(\omega)=-\varepsilon_d$ ($\varepsilon_\perp(\omega)=-1$ for h-BN surrounded by air). It is the typical dispersion limit of surface polaritons.[27] When changing the environment from air to a dielectric material with $\varepsilon_d$=3, the SO frequency shifts to $\omega$=1515 cm$^{-1}$, where $\varepsilon_\perp$= −3. The dependence of SHP properties on the environment (similar to that surface polaritons on isotropic materials) promises sensing applications as well as active HSP control based on switching the dielectric envrioment.[29]

h-BN and other 2D vdW samples are typically obtained by exfoliation, yielding thin flakes (slabs), where the OA is oriented perpendicular to the horizontal flake surface. According to Fig. 1a, a dipole placed above the flake cannot excite HSPs. However, HSPs can be excited at the edges of thin flakes (sketch in Fig. 2a), when the dipole is placed directly above an edge. The numerical simulation in Fig. 2b clearly shows dipole-launched HSP rays on the edge of a h-BN flake, propagating away from the dipole. Multiple reflections at the top and bottom corners of the edge yield a zig-zag field pattern along the edge. Simultaneously, the dipole launches HPs. They reflect at the bottom and top surfaces of the flake, yielding bright half-circles (corresponding to strong electromagnetic field concentration) on the flake surfaces (marked by dashed green half-circles on the top surface). The distance ($d_1$~180 nm) between two consecutive field maxima along the upper corner of the edge is about 20% reduced compared to the distance ($d_2$~215 nm) between two consecutive bright fringes on the top surface, which can be explained by the different propagation angles of the HSP and HP rays, respectively ($\theta_v-\theta_s$=5° at 1535 cm$^{-1}$).

The zig-zag HSP ray forms a guided polariton that can be described by a superposition of eigenmodes (labeled as SMn (n = 0,1,2…). Their profiles are depicted in Fig. 2c, clearly showing the mode confinement to the edge of the h-BN slab, which becomes stronger with larger n. The eigenmodes are described by a complex-valued wavevector $K_{SMn}=q_{SMn}+i\gamma_{SMn}$.[10] The propagation constants $q_{SMn}$ relate to the mode wavelengths as $\lambda_{SMn}=2\pi/q_{SMn}$, while the damping parameter $\gamma_{SMn}$ determines the mode propagation length $L_{SMn}=1/\gamma_{SMn}$. The largest propagation length is found for the SM0 mode, as its wavevector is smaller than that of the higher order modes.

We experimentally verify HSPs with nanoscale-resolved infrared images (Fig. 3a) taken with a scattering-type scanning near-field optical microscope (s-SNOM, Neaspec)[5,6,8-10]. As a sample we used a 40-nm-thick h-BN flake on a Si/SiO$_2$(250 nm) substrate, following the exfoliation procedure described in the Methods. As illustrated in Fig. 3b (upper panel), the sharp metallized tip of the s-SNOM (tip radius~30 nm) was illuminated by a wavelength-tunable mid-infrared quantum cascade laser (operating between 1295 − 1440 cm$^{-1}$). Acting as an antenna, the tip converts the incident light into strongly enhanced and confined near fields at the very tip apex.[5,6,8-10] Similar to the dipole source in Figs. 1 and 2, they provide the necessary wavevectors for launching HPs and HSPs.

We illustrate the tip-launching of HSPs with a numerical simulation shown in Fig. 3b (displaying the in-plane electrical field component $|E_x|$). Similar to Fig. 2, we find tip-launched HSP rays (zig-zag pattern) on the edge of the h-BN flake. The SM0 mode – exhibiting the largest wavelength and propagation length - interferes with the incident field, yielding interference fringes (indicated by blue arrows in the simulation of Fig. 3b) along the edge.

Figure 3a shows s-SNOM images of a h-BN flake for three different frequencies (1440 cm$^{-1}$, 1420 cm$^{-1}$ and 1410 cm$^{-1}$). Inside the flake, we clearly see interference fringes parallel to the edges of the flake, which have been already observed and analyzed in previous publications.[5,10] They result from the interference of the local field below the tip with the tip-launched and edge-reflected fundamental HP waveguide mode (labeled M0).[5,10] The fringe spacing consequently corresponds to half of the M0 mode wavelength, $\lambda_{M0}/2$ (indicated in Fig. 3a, middle image). Most intriguingly, our images reveal, for the first time, signal oscillations (manifesting as bright and black dots along the edges) that are strongly confined to the flake edges, which exhibit a reduced oscillation period. We attribute them to the excitation of the SM0 mode, as predicted in Fig. 3b. The tip`s near field launches hyperbolic surface modes that propagate along the edge of the flake. They reflect at the flake termination (bottom left in Fig. 3a) and interfere with the field below the tip apex. From the oscillation period at the edges ($\lambda_{SM0}/2$, indicated in the middle image of Fig. 3a) we can thus directly measure the wavelength of the SM0 mode. We also recognize that the oscillations decay with distance to the flake termination, analogous to the decay of the fringes inside the flake. We can understand this observation by the finite propagation length of the SM0 mode. Note that the images of Fig. 3a strongly resemble s-SNOM images of graphene nanostructures[30,31], where signal oscillations inside and at the edges of the graphene nanostructures have been explained by the interference of tip-launched sheet and edge plasmons, respectively.

For measuring the dispersion $q(\omega)$ and propagation lengths $L(\omega)$ of the SM0 mode and comparing them to those of the M0 mode, we extracted and fitted s-SNOM line profiles along and parallel to the horizontal flake edge, at the positions marked by the blue and red arrows in Fig. 3a. They are shown in Fig. 3c for 1410 cm$^{-1}$ and 1440 cm$^{-1}$ (symbols) To measure the wavevector and the propagation length of the surface mode SM0, we fitted the line profiles along the edges (blue symbols in Fig. 3c) by a damped sine-wave function of the form $e^{-2\gamma_{SM0}x}\sin(2q_{SM0}x)$ (yellow curves in Fig. 3c). The line profiles inside the flake (blue symbols) were fitted by $e^{-2\gamma_{M0}x}\sin(2q_{M0}x)/\sqrt{2x}$ (yellow curves). The additional factor $\sqrt{2x}$ accounts for the radial propagation (i.e. decay) of the volume modes inside the flake.[32] To exclude disturbing effects such as multiple reflections between tip and flake termination, as well as the simultaneous probing of surface and volume modes, we excluded the first fringe of each profile from the fitting. Fig. 3c shows that the fitting results (yellow solid lines) agree well with the experimental line profiles (blue and red symbols).

Figure 4a shows the experimental and calculated dispersion of the fundamental hyperbolic volume and surface modes of the h-BN flake, $q_{M0}(\omega)$ and $q_{SM0}(\omega)$. The experimental data were obtained by repeated imaging of the flake and fitting of line profiles for different wavelength in the range between 1410 and 1440 cm$^{-1}$, analogous to Fig. 3c. Both the dispersion data for the M0 (blue open symbols) and SM0 (red open symbols) modes agree well with calculations (solid blue and red lines) for a 40 nm thick h-BN flake on a SiO$_2$ substrate obtained with mode analysis (finite-element method, see Methods). For a given frequency, the propagation constant of the SM0 mode is larger than that of the M0 mode, $q_{SM0}(\omega) > q_{M0}(\omega)$, indicating a higher electromagnetic field confinement. We note that for thinner h-BN flakes both $q_{SM0}$ and $q_{M0}$ are increased, while $q_{SM0}$ is again larger than $q_{M0}$, as observed in s-SNOM images of different flakes and by numerical calculations. This thickness dependence follows the general trend of polaritons to become more confined with decreasing structures size (e.g. thickness and width of polaritonic waveguides). In Fig. 4a we also observe that the simulated SM0 dispersion approaches asymptotically the SO frequency at $\omega$ = 1568 cm$^{-1}$ ($\varepsilon_\perp$=−1, flake edges surrounded by air), in contrast to the simulated M0 dispersion extending above the SO frequency[5-10]. We stress that the SO frequency is the typical dispersion limit for surface modes[25]. Our experimental and theoretical results thus verify that the measured SM0 mode indeed is a hyperbolic surface phonon polariton propagating along the edge of the h-BN flake. In Fig.4b we show the group velocities $v_g$ of the M0 and SM0 modes, obtained by calculating $v_g = \partial\omega/\partial q$. A good agreement between experimental (symbols) and calculated (solid lines) results are found. Consistent with a recent s-SNOM study[10], the M0 modes exhibit ultraslow group velocities. The SM0 group velocity is even slower. For example, at $\omega$ = 1440 cm$^{-1}$ we find $v_{g,SM0} \approx 0.005c$ compared to $v_{g,M0} \approx 0.007c$.

In Fig. 4a we also show the calculated dispersion of the M0 and SM0 modes of a 40-nm-thick free-standing h-BN flake (dashed blue and dashed red lines). We find that both curves are shifted to smaller wavevectors, as compared to the results with the substrate. We explain the larger wavevectors in case of the substrate by the penetration of the polariton fields into the substrate. Because of the higher refractive index of the Si/SiO$_2$ substrate compared to air, the effective mode index of the hyperbolic polaritons is increased. Interestingly, the substrate induces a larger wavevector shift for the M0 mode than for the SM0 mode. As the substrate is not in direct contact to the surface of the h-BN edge, the fields of the SM0 penetrate less into the substrate than those of the M0 mode (see calculated cross-sections of the SM0 and M0 mode profiles in Figs. 2c, S1a and S1b). Consequently, the SM0 mode is less affected by the substrate than the M0 mode.

To estimate the confinement of the SM0 mode into the flake (along the top h-BN surface), we analyzed the s-SNOM image at 1410 cm$^{-1}$. By plotting in Fig. 3d various line profiles parallel to the edge, we find that the oscillation period $p$ is half of the wavelength of the surface mode, $p(y) = \lambda_{SM0}/2$, for $y$ < 1000 nm, while for $y$ > 1000 nm we find $p(y) = \lambda_{M0}/2$. From this observation we conclude that the surface mode SM0 extends less than 1000 nm into the flake, which is in good agreement with the 1/$e$ field decay length (about 700 nm) of the calculated SM0 mode profile at 1410 cm$^{-1}$. Because of the complex interference pattern formed by the M0 and SM0 modes propagating in both *x*- and *y*-directions, a more quantitative analysis will require further studies beyond this work.

To study the propagating length and lifetime of the M0 and SM0 modes, we plot $L_{M0}$ = $1/\gamma_{M0}$ and $L_{SM0}$ = $1/\gamma_{SM0}$ in Fig. 5a as a function of frequency. The experimental data (blue and red open symbols, respectively) were obtained by fitting of the s-SNOM line profiles as described above. We observe that the propagation lengths of both SM0 and M0 modes decrease with increasing frequency. Moreover, the M0 mode propagates farther than the SM0 mode, as confirmed by calculations (see Methods).

Although the SM0 mode has a shorter propagation length compared to the M0 mode, it does not necessarily imply a smaller lifetime, as the SM0 modes have a smaller group velocity (Fig. 4b). To elucidate this interesting issue, we plot in Fig. 5b the lifetimes of both modes as a function of frequency. To that end, we calculated the lifetimes according to $\tau_{SM0}$ = $L_{SM0}/2v_{g,SM0}$ and $\tau_{M0}$ = $L_{M0}/2v_{g,M0}$. The experimental lifetimes (symbols in Fig. 5b) amounting to about 0.8 ps were obtained from the propagation lengths of Fig. 5a (symbols) and group velocities of Fig.4b (symbols). Interestingly, the lifetimes for M0 (blue symbols in Fig.5b) and SM0 (red symbols) modes are nearly the same, and only slightly vary with frequency. These experimental observations are confirmed by the lifetimes (solid lines in Fig.5b) calculated from the simulated propagation lengths and group velocities.

We note that the measured lifetimes are not an intrinsic property of the SM0 and M0 modes of h-BN flakes, as they are modified by the frequency-dispersive and lossy Si/SiO$_2$ substrate (as discussed in Fig. 4a). For that reason, we also calculated the lifetimes of the SM0 and M0 modes of a 40-nm-thick free-standing h-BN flake (dash-dotted lines in Fig. 5b). In this case, the lifetimes of both modes are significantly increased to around 2-3 ps. Interestingly, the lifetime of the M0 mode is constant with frequency, while the lifetime of the SM0 mode significantly decreases with increasing frequency. We attribute these distinct trends to the different field confinement of the SM0 and M0 modes. Approaching the SO frequency, the confinement of the SM0 to the surface increases, which results in a larger fraction of polariton field propagating inside the h-BN flake. As damping of the surface polaritons is caused by dissipation of field energy inside the material (i.e. inside the h-BN flake), the lifetime of the SM0 modes reaches its minimum at the SO frequency. The constant lifetime of the M0 mode we attribute to the volume confinement of the fields of HPs. Independent of the frequency, the polariton field propagates inside the h-BN flake. Certainly, further studies will be needed in the future for a more comprehensive understanding of this observation.

We finally note that in the images and line profiles of Fig. 3c we do not observe higher order SMn modes (n > 0), although in principle they are excited by the s-SNOM tip (leading to the zig-zag pattern in the numerical simulation shown in Fig. 3b), similar to the observation of higher-order Mn modes[9]. However, their wavelengths are significantly smaller than that of the fundamental SM0 mode (see the calculated dispersion in Figure S1c of the Supporting Information), which makes them challenging to be resolved experimentally. They might be observed in future s-SNOM experiments by using h-BN flakes with sharper edges, reducing the imaging pixel size and improving the signal-to-noise ratio.

In summary, we employed spectroscopic s-SNOM for real-space nanoimaging of hyperbolic surface phonon polariton modes that are confined and guided at the edges of thin h-BN flakes on a Si/SiO$_2$ substrate. From the near-field images we extracted the dispersion and the propagation length of the fundamental HSP-M0 waveguide mode, and compared them to the fundamental HP-M0 waveguide mode measured on the same flake. We found that the HSP-SM0 mode exhibits stronger field confinement (i.e. higher mode momenta), smaller group velocities and nearly identically long lifetimes. Our results provide most fundamental insights for the development of compact mid-infrared photonic devices based on hyperbolic surface and volume polaritons, such as metamaterials[33], resonators[7], waveguides[9] and imaging devices[34].

**Methods**

*Sample preparation.* High quality, homogeneous and large area h-BN flakes were transferred onto a Si/SiO$_2$ (250nm) substrate. To that end, we first performed mechanical exfoliation of commercially available h-BN crystals (HQ graphene Co, N2A1) using a blue Nitto tape (Nitto Denko Co., SPV 224P). Then, from the h-BN flakes attached to the tape, we performed a second exfoliation step onto several Polydimethyl-siloxane (PDMS) transparent stamps. Via optical inspection of the flakes transferred to the stamps using an optical microscope, and atomic force microscopy characterization (AFM) of them, large flakes of desired thickness and with well-defined corners were identified. Those flakes were finally transferred onto different places of the Si/SiO$_2$ substrate by means of the deterministic dry transfer technique[35].

*Near-field optical nanoimaging.* Our s-SNOM is a commercial system (from Neaspec GmbH) based on an atomic force microscope (AFM). The metallized AFM tip oscillates vertically with an amplitude of about 50 nm at a frequency $\Omega \approx$ 270 kHz. It is illuminated by light from a wavelength-tunable continuous-wave quantum cascade laser (QCL). The backscattered light is collected with a pseudo-heterodyne interferometer[36]. To suppress background contribution in the tip-scattered field, the interferometric detector signal is demodulated at a higher harmonic n$\Omega$ ($n \geq 2$), yielding near-field amplitude $s_n$ and phase $\varphi_n$ images. Fig. 3a shows amplitude $s_3$ images.

*Numerical simulations.* The simulation results shown in Figs. 4 and 5 were obtained using the mode solver of commercial software package COMSOL. We considered a 40-nm-thick h-BN flake on a Si/SiO$_2$(250nm) substrate. The SiO$_2$ dielectric function was taken from ref. 37. The Si permittivity was chosen to be $\varepsilon_{Si} = 12$, for all frequencies constant[37]. Both axial and transverse components of the h-BN permittivity were described by a Lorentz model according to [5-10]

$$\varepsilon_{\perp,\parallel}(\omega) = \varepsilon_\infty \left(1 + \frac{\omega_{LO}^2 - \omega_{TO}^2}{\omega_{TO}^2 - \omega^2 - i\omega\Gamma}\right)$$

where $\omega_{LO}$ and $\omega_{TO}$ are the TO and LO phonon frequencies, $\Gamma$ the damping constant and $\epsilon_\infty$ is the high-frequency permittivity. For the transverse components we used $\varepsilon_{,\infty} = 4.9$, $\omega_{,LO} = 1610\ cm^{-1}$, $\omega_{,TO} = 1360\ cm^{-1}$. For the axial component we used $\varepsilon_{,\infty} = 4.95$, $\omega_{,LO} = 825\ cm^{-1}$, $\omega_{,TO} = 760\ cm^{-1}$. In order to fit the experimental lifetimes in Fig. 4b, we used the damping constant $\Gamma = 2\ cm^{-1}$ for both components. A similar value of the damping constant is reported in ref. 7.

**Acknowledgement**

The authors acknowledge support from the European Union through ERC starting grants (TERATOMO grant no. 258461 and SPINTROS grant no. 257654), the European Commission under the Graphene Flagship (GrapheneCore1, grant no. 696656), and the Spanish Ministry of Econoy and Competitiveness (national projects MAT2014-53432-C5-4-R, MAT2012-36580, MAT2015-65525-R, MAT2012-37638 and MAT2015-65159-R).

**Competing Finantial Interests**

R.H is co-founder of Neaspec GmbH, a company producing scattering-type scanning near-field optical microscope systems, such as the one used in this study. The remaining authors declare no competing financial interests.

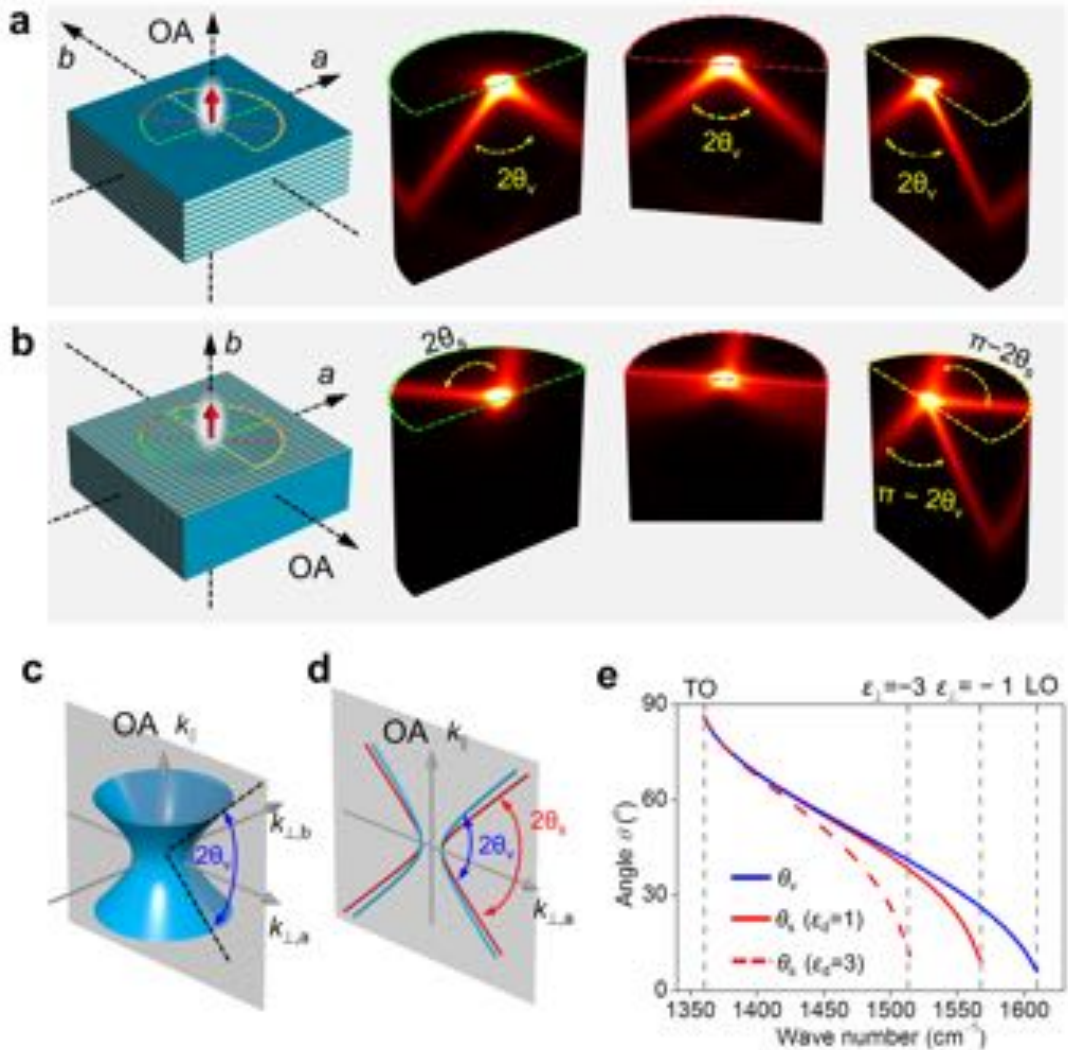

**Figure 1 Dipole-launching of HSPs and HPs in bulk h-BN. a**, Left, sketch of dipole on bulk h-BN with optical axis (OA, indicated by a black arrow) perpendicular to the atomic layers. Right, different views of the numerically calculated electric-field intensity distribution inside the h-BN, obtained by cutting out the volume marked by dashed half-circles in the sketch to the left. The bright rays reveal the dipole-launched HPs that propagate at an angle $\theta_v$ with respect to the OA. **b**, Left, sketch of dipole on bulk h-BN with OA parallel to the top horizontal surface. Right, different views of the numerically calculated electric-field intensity distribution inside the h-BN, obtained by cutting out the marked by dashed half-circles in the sketch to the left. The bright rays inside the h-BN reveal the dipole-launched HPs, while the bright rays along the surface reveal the dipole-launched HSPs (with an angle $\theta_s$ with respect to the OA). **c**, Isofrequency surface of HP modes in the h-BN volume. **d**, Comparison of the HSP isofrequency curve (red) with a 2D cross-section of the HP isofrequency surface (blue curve). **e**, Propagation angles $\theta_s$ and $\theta_v$ as a function of frequency, for h-BN environments of a permittivity $\varepsilon_d$.

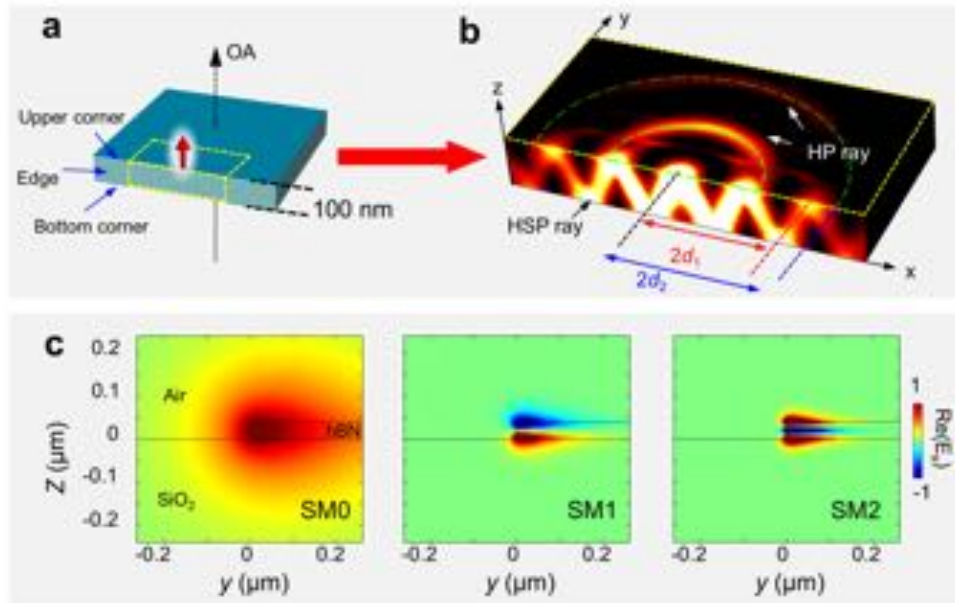

**Figure 2 Dipole-launched HSPs at the edge of a thin h-BN flake**. **a**, Sketch of dipole at the edge of a 100-nm-thick h-BN flake. The OA of the h-BN flake is perpendicular to the atomic layers. **b**, Simulated electric-field intensity distribution corresponding to the sketch in Fig. 2a. The zig-zag rays reveal HSPs being reflected multiple times at the corners of the flake edge. The bright half circles reveal HP rays being reflected multiple times at the flake surfaces. **c**, Profiles of the first three HSP modes SMn (n = 0,1,2) at the edge of a 40-nm-thick h-BN flake on the SiO$_2$ substrate at 1440 cm$^{-1}$.

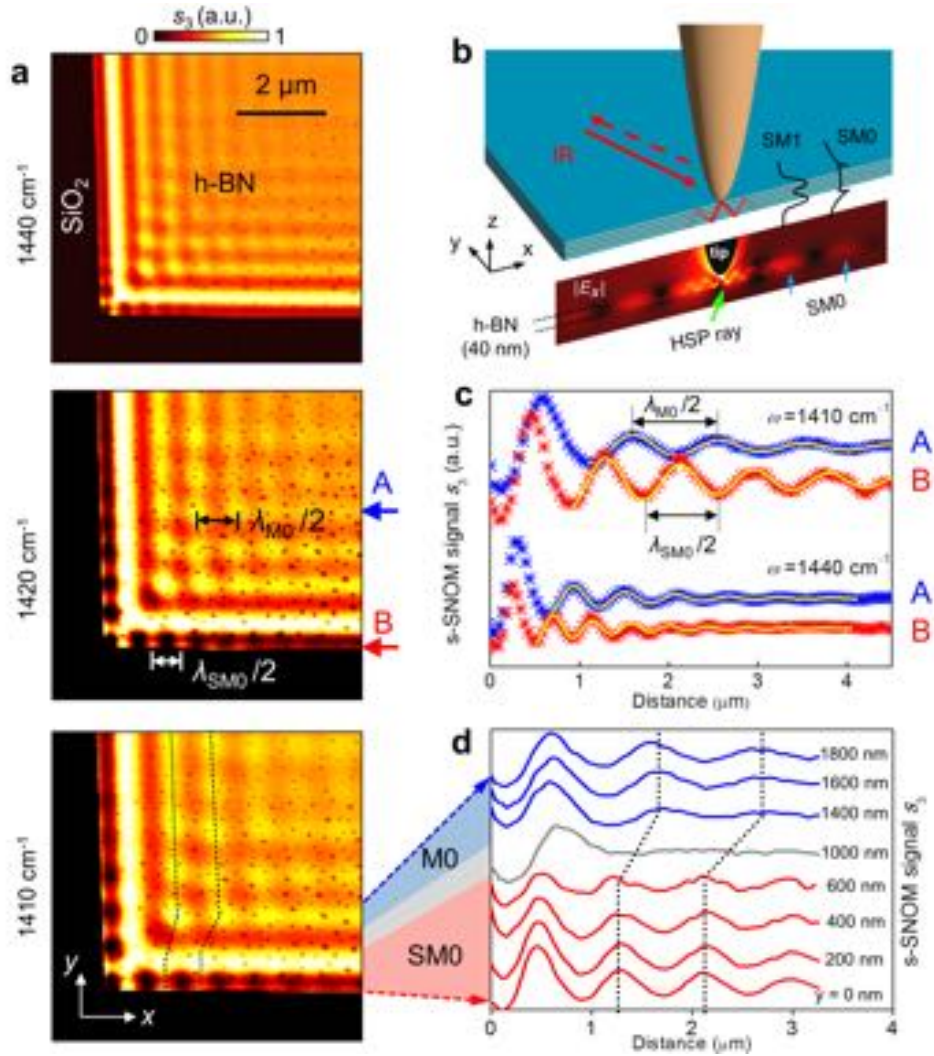

**Figure 3 Near-field imaging of HPs and HSPs of a 40-nm-thick h-BN flake**. **a**, Experimental infrared s-SNOM images (amplitude signal $s_3$) at three different frequencies. The black and white arrows in the image at 1420 cm$^{-1}$ indicate the periods of near-field oscillations on the h-BN flake and its edge, corresponding to half the wavelengths of the M0 and SM0 modes, respectively. Small black dots on the h-BN surface are due to sample contamination. **b**, Illustration of the experiment. The s-SNOM tip is illuminated (solid red arrow) with mid-IR light and launches HSPs at the edge of a h-BN flake with OA perpendicular to the atomic layers. The backscattered field (dashed red arrow) is recorded as function of tip position. Below the sketch, the launching of HSPs (zig-zag rays, indicated by a green arrow) is demonstrated by a numerical simulation of the electric field distribution around tip and h-BN flake. **c**, Horizontal line profiles extracted from Fig. 3a at the positions marked by blue and red arrows. The yellow curves display fittings to the experimental data as described in the main text. **d**, Horizontal line profiles extracted from the s-SNOM image at 1410 cm$^{-1}$ at different vertical positions *y*. All curves are normalized to their maximum value.

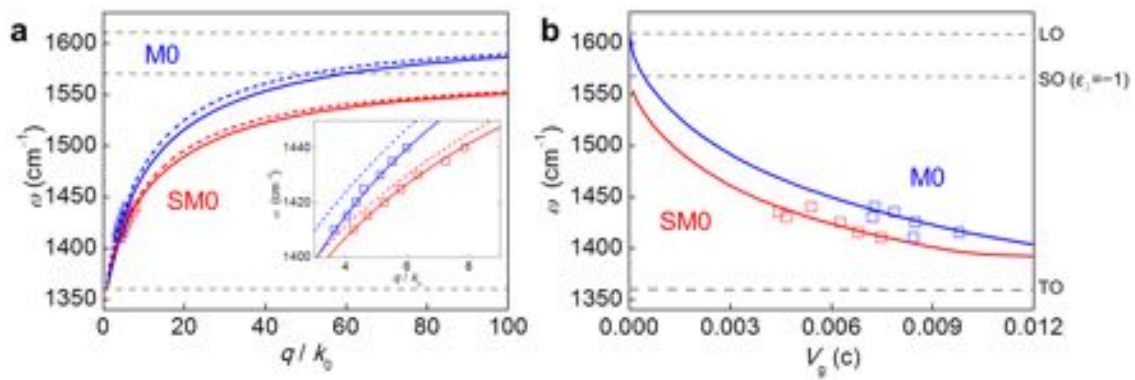

**Figure 4 Polariton dispersion and group velocities. a**, Experimental (symbols) and calculated (solid lines) polariton dispersions of SM0 (red) and M0 (blue) modes of a 40 nm thick h-BN flake on a $SiO_2$ on Si substrate. Dashed lines are the calculated dispersion of SM0 (red) and M0 (blue) modes of a 40 nm thick free-standing h-BN flake (without any substrate). The experimental data were obtained by fitting s-SNOM lines profiles such as the ones shown in Fig. 3c. $k_0$ is the wavevector in free space. **b**, Experimental (symbols) and calculated (solid lines) group velocities of SM0 (red) and M0 (blue) modes. They were obtained from the polariton dispersion according to $v_g = \partial\omega/\partial q$.

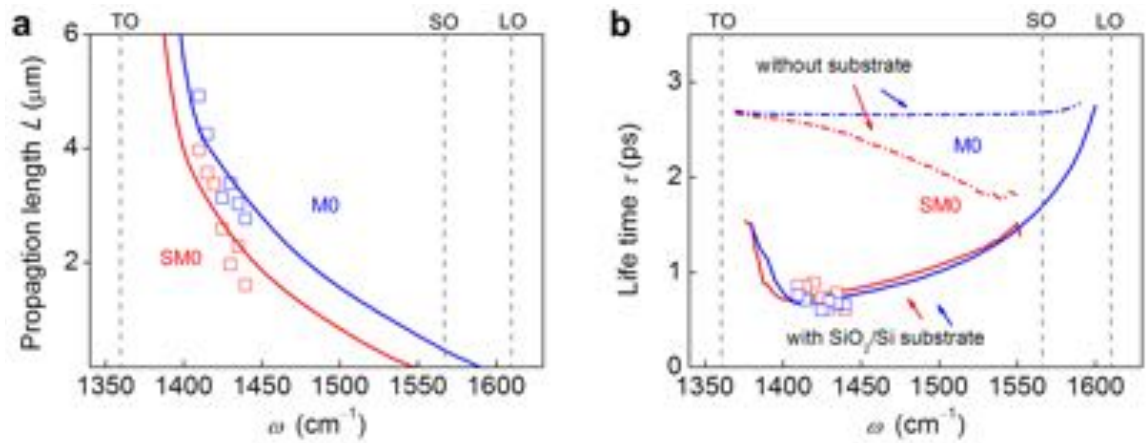

**Figure 5 Propagation lengths and lifetimes. a**, Experimental (symbols) and calculated (solid lines) propagation lengths of SM0 (red) and M0 (blue) modes. The experimental data were obtained by fitting s-SNOM lines profiles such as the ones shown in Fig. 3c. **b**, Experimental (symbols) and calculated (solid lines) lifetimes of SM0 (red) and M0 (blue) modes. For comparison, the dashed-dotted lines show the calculated lifetimes of SM0 (red) and M0 (blue) modes for an h-BN flake without substrate.

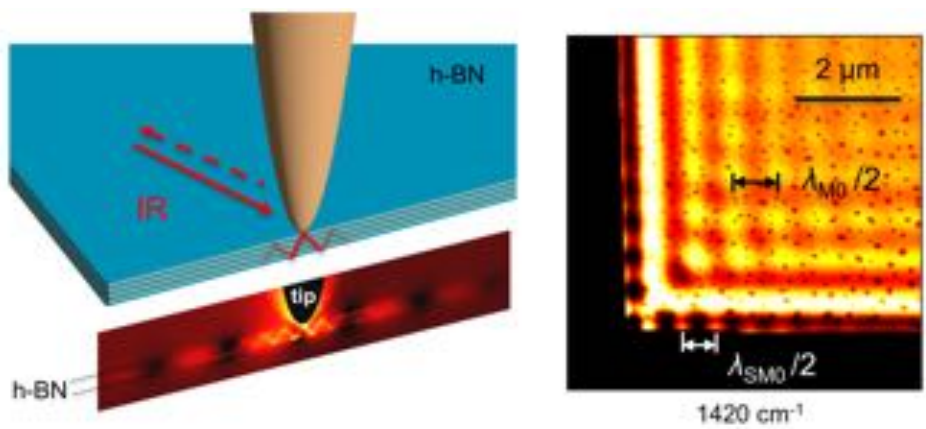

TOC graphic